\documentclass[useAMS,usenatbib]{mn2e}
\usepackage{graphicx}
 \usepackage{epstopdf}
 \usepackage{amsmath}
\usepackage{color}

\title[QLD in SgrA] {Synchrotron emission from a nearby zone of SgrA$^*$}

\author[Gogaberishvili et al.]{G. Gogaberishvili,$^{1}$
Z.N. Osmanov$^{1,2}$\thanks{E-mail: z.osmanov@freeuni.edu.ge} \& S.M. Mahajan$^{3}$\\
$^{1}$School of Physics, Free University of Tbilisi, 0183, Tbilisi, Georgia\\
$^{2}$ E. Kharadze Georgian National Astrophysical Observatory, Abastumani, 0301, Georgia.\\
$^{3}$ Institute for Fusion Studies, The University of Texas at
Austin, Austin, TX 78712, USA\\
}

\begin{document}

\pagerange{\pageref{firstpage}--\pageref{lastpage}} \pubyear{2016}

\maketitle

\label{firstpage}

\begin{abstract}
Quasi-linear diffusion (QLD), driven by the cyclotron instability, is proposed as a mechanism for the possible generation of synchrotron emission in the nearby zone of SgrA$^*$. For physically reasonable parameters, the QLD, by causing non-zero pitch angle scattering lets electrons with the relativistic factors of the order of $10^8$ emit synchrotron  radiation in the hard $X$-ray spectral band $\sim120$ keV.

\end{abstract}


\section{Introduction}
Very high energy (VHE) astronomy took some giant strides in the last decade. The diffuse VHE electromagnetic radiation from the central black hole of the Milky Way, observed by the High Energy Stereoscopic System (HESS) \citep{pevhess}collaboration, detected $\gamma$-rays with a photon index of $\sim 2.3$ and energies extending to the TEV range. Such VHE photons  are likely to be generated via the hadronic pp channels. This discovery has opened a door for an additional study of a mechanism responsible for generation of VHE particles, their sources like  Sagittarius A$^*$ (Sgr A$^*$), and the consequent radiation. It is clear that energetic particles/emission pump energy from the preponderant gravitational field, but pinpointing of a specific mechanism is, still, a matter of discussion; some possible contenders are:  models of diffusive acceleration \citep{stoch}, the Langmuir-Landau-centrifugal-drive \citep{srep,pev}, and magnetospheric gap acceleration \citep{rieger}.

Recent studies have established that the $X$-ray band \citep{xr,xray,rosat} in the radiation emission from Sgr A$^*$ might have synchrotron origin. Therefore, let us examine the role of synchrotron emission in the generation of $X$-rays. Because the magnetic field close to the central black hole (BH) is strong, synchrotron emission from electrons is expected to be very efficient. Consequently, the particles, very rapidly, lose their transversal momentum (component perpendicular to the magnetic field line), fall to the ground Landau level terminating the emission. 

Let us estimate the expected time for the duration of emission. Remembering the synchrotron emission power of a single relativistic electron, $P_s\simeq 2e^4B^2\gamma^2/(3m^2c^3)$, one can show that for physical parameters typical for the magnetosphere of Sgr A$^*$ (including $B\sim 10$G \citep{pev}), and the corresponding cooling time-scale $t_s=\gamma mc^2/P_s$, the electrons with a relativistic factor $\gamma = 10^6$ will cease to radiate in $\sim 2.6$ sec. 

Compare it to the kinematic time-scale of  $\sim190$ sec (two orders of magnitude larger); the latter being the rotation period of the central BH $P = 2\pi/\Omega$, where $\Omega\simeq ac^3/(GM)$ is the angular velocity of the nearby area of the BH. Here we have used the following parameters of Sgr A$^*$: $a\simeq 0.65$ \citep{a} and $M\simeq 4\times 10^6\; M_{\odot}$ \citep{M} ($M_{\odot}$ is the Solar mass). 

One must further note \cite{stoch,pev,rieger} that for (much) higher particle energies, the synchrotron cooling time-scale will be even shorter. One has to emphasize that considerably higher Lorentz factors $\sim 10^{10}$ are possible, for example, in an equilibrium scenario where  the (centrifugal) acceleration timescale, $t_{acc}\simeq \frac{1}{2}\Omega^{-1}\gamma^{-1/2}$ is of the order of $t_s$. Similar situation could pertain in a stochastic acceleration process \citep{stoch}, and in the gap-type acceleration mechanism \citep{rieger}, indicating that the perpendicular momentum via the synchrotron losses vanishes very rapidly.

Preceding discussion clearly indicates that the standard synchrotron mechanism can be significant only in relatively distant regions from the BH (see the paragraph following Eq. (\ref{B})).  Of course, if there were to exist a mechanism that could refurbish perpendicular energy, the particle could continue emitting  synchrotron radiation.  One such mechanism, invoking the cyclotron instability driven by the anomalous Doppler effect [\cite{kmm}], demonstrated that the instability driven quasi-linear diffusion could push particles across the magnetic field lines. As a result, non zero pitch angles are preserved and the corresponding synchrotron mechanism is maintained. This mechanism has been applied to active galactic nuclei \citep{agn}, pulsars \citep{puls} and magnetars \citep{mag} and it was shown that the quasi linear diffusion (QLD) might be a significant mechanism maintaining a continuous synchrotron emission.

The present paper, investigating the dynamics of relativistic electrons, aims to show how synchrotron emission is maintained via  QLD in the magnetosphere of Sgr A$^*$.

The paper is organized as follows: in Sec.2 we introduce the theoretical model, in Sec. 3 we apply the model to Sgr A$^*$ and derive the relevant results,  and in Sec. 4 we discuss and summarize them

\section{Theoretical model}
In the rotating magnetospheres of compact objects, the electron-positron plasmas may be viewed as consisting of two components:  the bulk with  relatively small Lorentz  factors, $\gamma_{p}$, and a smaller high $\gamma_{b}$ beam component \cite{LMU,MU}. This general description, naturally, applies to the class of objects like the nearby zone of SgrA$^*$ \citep{agn}. Close to BHs, the accretion matter has a very high temperature resulting in full ionization of the accretion flow creating a plasma. In the same region, efficient pair creation may take place \citep{pair} potentially leading to the formation of an electron-positron plasma.

It was shown in \cite{review} the length-scale of the acceleration region is  $\sim R_{lc}/\gamma_b$, where  $R_{lc} = c/\Omega$ is the radius of the light cylinder (LC) - a hypothetical zone, where the linear velocity of rotation equals the speed of light and $\gamma_b\sim 10^8$. The resulting acceleration shell is very thin - one can, therefore,the  assume that energy in this area is almost uniformly distributed among different species: $n_{b}\gamma_{b}\simeq n_{p}\gamma_{p}$. It is worth noting that, unlike the synchrotron mechanism that does not impose any significant constraints on particle dynamics, other restricting factors will lead to the equilibrium between acceleration and energy losses. It has been shown by \cite{kmm}  that in the regime of the frozen-in condition the plasma is subject to the anomalous Doppler effect, which induces the unstable resonance cyclotron modes 
\begin{equation}\label{res}
\omega - k_{_{\|}}c-k_xu_x-\frac{\omega_B}{\gamma_{b}} = 0
\end{equation}
where $k_{\parallel}$ denotes the longitudinal (along the magnetic field lines) component of the wave vector, $k_{x}$ is the component along the drift, $u_x \approx c^2\gamma_{b}/{\rho\omega_B} $ denotes the curvature drift velocity, c is the speed of light, $\rho$ is the curvature radius of the magnetic field lines and $\omega_B={eB}/{mc}$ is the cyclotron frequency.

The cyclotron frequency corresponding to the resonance condition is given by \citep{MM}:

\begin{equation}\label{nu}
\nu \approx \frac{\omega_{_B}}{2\pi\delta\gamma_b},\;\;\;\;\; \delta
= \frac{\omega_p^2}{4\omega_{_B}^2\gamma_p^3},
\end{equation}
where $\omega_{p}=\sqrt {4\pi{n}_{p}e^2/m}$ denotes the Langmuir frequency of the electron-positron plasma component; $n_{p}$ is the corresponding number density. 

Outside the LC zone, the  dynamics is predominantly governed by accretion \cite{pev}. Let us further assume that some fraction $\eta<1$ of the whole kinetic energy of accretion matter is transferred to emission. One could, then, estimate  the number density by the following expression \citep{pev}
\begin{equation}\label{np}
n_{p}=\frac{L}{4\eta\pi{m}_{p}c^2v{R}_{lc}^2},
\end{equation}
where $m_p$ is the proton's mass and
\begin{equation}\label{np}
v=c\;\sqrt{1-\left(\frac{c^2}{c^2+\frac{GM}{R_{lc}}}\right)^2}
\end{equation}
is the velocity of the accreting matter near the LC. Throughout the paper we use $\eta = 0.1$. It is worth noting that the possible maximum efficiency of energy conversion in the accretion process is of the order of 0.25 \citep{carroll}).

To study QLD, one should take into account two forces that take part in the diffusive distribution process. 
.Following \citep{landau}, one of the dissipative forces is the synchrotron radiation reaction force, 
\begin{equation}\label{fs}
    F_{\perp}=-\alpha_{s}\frac{p_{\perp}}{p_{\parallel}}\left(1+\frac{p_{\perp}^{2}}{m^{2}c^{2}}\right),
    F_{\parallel}=-\frac{\alpha_{s}}{m^{2}c^{2}}p_{\perp}^{2},
\end{equation}
where $\alpha_{s}=2e^{2}\omega_{_B}^{2}/3c^{2}$ and $p_{\parallel}$ and $p_{\perp}$ 
are the longitudinal and transversal components of the momentum. 

The second force is originated in the non-uniformity of the magnetic field. In particular, one can show that if the magnetic field lines are curved, the particles experience the force with the following components \citep{landau}
\begin{equation}\label{g}
G_{\perp} = -\frac{cp_{\perp}}{\rho},\;\;\;\;\;G_{_{\|}} =
\frac{cp_{\perp}^2}{\rho p_{\parallel}},
\end{equation}
where $\rho$ is the curvature radius of the magnetic field lines.

Under the action of these forces, the kinetic equation for the particle distribution function, $f$, may be written as \citep{LMU}
\begin{align} \label{qld1}
    \frac{\partial\textit{f }}{\partial
    t}+\frac{1}{p_{\perp}}\frac{\partial}{\partial p_{\perp}}\left(p_{\perp}
    \left[F_{\perp}+G_{\perp}\right]\textit{f }\right)=\nonumber \\
    =\frac{1}{p_{\perp}}\frac{\partial}{\partial p_{\perp}}\left(p_{\perp}
D_{\perp,\perp}\frac{\partial\textit{f }}{\partial
p_{\perp}}\right),~~~
\end{align}
where $D_{\perp,\perp}={D}\delta{E}_{k}^2$ denotes the diffusion coefficient, $D={e^2}/{8c}$ \citep{puls} and $|E_{k}|^2$ is the corresponding wave energy density per unit of wavelength. With the plausible assumption that half of the beam's energy density, $mc^2n_{b}\gamma_{b}^2$, converts to that of the waves, $|E_{k}|^2k$ \citep{MU,LMU} one obtains
\begin{equation}\label{ek2}
|E_k|^2 = \frac{mc^3n_b\gamma_b} {4\pi\nu}.
\end{equation}
If the energy were equipartitioned between the plasma and the beam, $n_b\simeq n_p\gamma_p/\gamma_b$. 

We next estimate the magnitude of the magnetic field by equating the magnetic energy density, $B^2/8\pi$ and the emission energy density, $L/4\pi r^2c$,  
$$B\simeq\sqrt{\frac{2L}{r^2c}}\simeq$$
\begin{align}\label{B}
 \simeq 27.5\times\frac{R_{lc}}{r}\times\left({\frac{L}{10^{37} \;ergs\;{s^{-1}}}}\right)^{1/2}\textup{G};
\end{align}
 it is a continuously increasing function of the bolometric luminosity, but this dependence is not very sensitive. It is worth noting that the synchrotron mechanism might become significant on distances for which the cooling timescale is large compared to the kinematic timescale, $r/c$; this takes place on distances larger than $\sim 50 R_{lc}$, where $B\leq 0.6$ G.

Due to the very efficient synchrotron cooling process, the pitch angles are, usually, very small. By combining Eqs. (\ref{fs},\ref{g},\ref{B}) one can make a straightforward estimate for the following ratio for physically realistic parameters
\begin{align}\label{pitc}
\frac{F_{\perp}}{G_{\perp}}\simeq 4.5\times10^{-7}\times\frac{L}{10^{37}{ergs\;{s^{-1}}}}\times \nonumber
 \\
\times\frac{\rho}{R_{lc}}\times\left(\frac{{R_{lc}}}{r}\right)^2\times\left(\frac{\psi}{10^{-5}\;rad}\right)^2<<1.
\end{align}
From Eq. (\ref{pitc}) it is clear that this condition is valid for the pitch angles satisfying $\psi\ll 0.015\;rad$. Therefore, $F_{\perp}$ can be neglected compared to $G_{\perp}$ in Eq. (\ref{qld1}) which, for the time-stationary case, $\partial/\partial t = 0$, reduces to 
\begin{align} \label{qld}
   \frac{1}{p_{\perp}}\frac{\partial}{\partial p_{\perp}}\left(p_{\perp}
    G_{\perp}\textit{f }\right)=\frac{1}{p_{\perp}}\frac{\partial}{\partial p_{\perp}}\left(p_{\perp}
D_{\perp,\perp}\frac{\partial\textit{f }}{\partial
p_{\perp}}\right)
\end{align}
 leading to the solution 
 \begin{equation}\label{f}
    \textit{f}(p_{\perp})=C exp\left(\int \frac{G_{\perp}}{D_{\perp,\perp}}
    dp_{\perp}\right)=Ce^{-\left(\frac{p_{\perp}}{p_{\perp_{0}}}\right)^{2}},
\end{equation}
where $C={\it const}$ and
\begin{equation}\label{p0}
     p_{\perp_{0}}\equiv\left(\frac{2\rho D_{\perp,\perp}}{c}\right)^{1/2}.
\end{equation}
Hence the average value of pitch angle is given by
\begin{equation}\label{pitch}
\bar{\psi}
 = \frac{1}{p_{\parallel}}\frac{\int_{0}^{\infty}p_{\perp} \textit{f}(p_{\perp})dp_{\perp}}{\int_{0}^{\infty}\textit{f}(p_{\perp})dp_{\perp}}
\approx \frac{1}{\sqrt{\pi}}\frac{p_{\perp_{0}}}{p_{\parallel}}.
\end{equation}
Due to the nonzero pitch angles, the relativistic particles radiate in the synchrotron regime emitting photons with energies \citep{RL}
\begin{equation}\label{eV}
\epsilon_{eV}\approx1.2\cdot10^{-8}B\gamma^2\sin \psi.
\end{equation}
%
The theoretical model we just established, is ready to be applied to SgrA$^*$; the agenda is to investigate the spectral characteristics of the emission process.

\section{Discusion}

The central black hole exhibits the characteristics of a very efficient accelerator-boosting proton energies up to PeVs. Through a series of well defined steps, these energies can be transferred to create relativistic electrons /electron-positron plasmas. That such processes may take place in the nearby zone of SgrA$^*$ was shown 
in a series of recent papers \citep{stoch,pev,rieger}. This paper, however, assumes  (without getting into the details of production) a strongly relativistic beam of electrons and investigates the associated synchrotron emission in the hard X-ray regime. One of the challenges was to find mechanisms that will allow a more or less continuous synchrotron radiation. We finally settled on to the quasilinear diffusion induced by a cyclotron instability. 

For ultra-relativistic electrons with the Lorentz factor of the order of $10^8$, one can estimate the frequency of the induced cyclotron waves (see Eq. (\ref{nu}))
\begin{align}
    \nu\simeq 6.9\times 10^3\times\frac{10^8}{\gamma_b}\times\left(\frac{\gamma_p}{2}\right)^3\times\nonumber
    \\
   \times\left(\frac{R_{lc}}{r}\right)^3 \times\left(\frac{L}{10^{37} ergs\; s^{-1}}\right)^{1/2}\textup{Hz}.
   \end{align}

\begin{figure}
  \resizebox{\hsize}{!}{\includegraphics[angle=0]{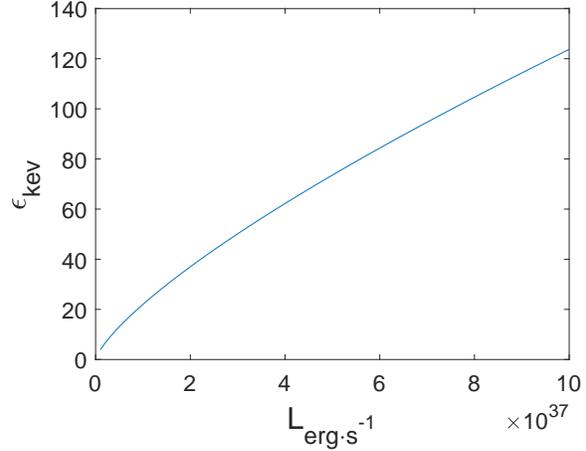}}
  \caption{Here we plot the emitted photon energy versus the bolometric luminosity. The set of parameters is: $\gamma_b = 10^8$, $\gamma_p = 2$, $a\simeq 0.65$ and $M\simeq 4\times 10^6\;M_{\odot}$. }\label{fig1}
\end{figure}

The frequency is, evidently, very sensitive to the the plasma $\gamma_p$.  The frequency increases with $\gamma_p$, but the pitch angle, on the other hand, is a continuously decreasing function of the relativistic factor,
\begin{align}\label{pitchangle}
\psi\simeq 6.7\times 10^{-6}\times\left(\frac{\gamma_{b}}{10^8}\right)^{1/2}\times\left(\frac{2}{\gamma_{p}}\right)^{5/2} \nonumber
    \\
    \times\left(\frac{L}{10^{37}ergs\; s^{-1}}\right)^{1/4}\times\left(\frac{r}{R_{lc}}\right)^{5/2}\textup{rad}.
\end{align}
For a large enough relativistic factor, then, the pitch angle ( measuring perpendicular energy) might become sufficiently small that the synchrotron emission may just shut off.

\begin{figure}
  \resizebox{\hsize}{!}{\includegraphics[angle=0]{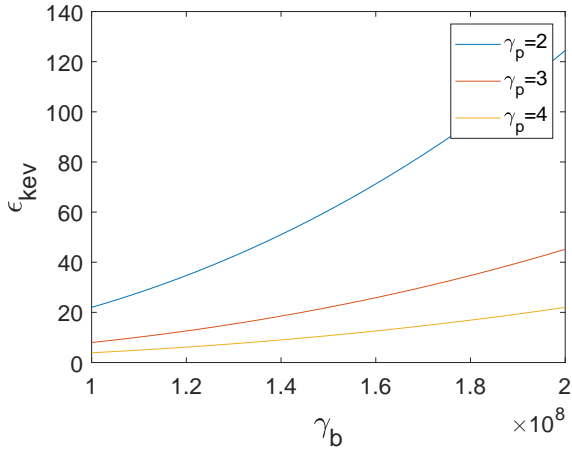}}
  \caption{Here we show the plots for $\epsilon_{_{keV}}(\gamma_b)$. The set of parameters is: $L = 10^{37}$ergs s$^{-1}$, $\gamma_p = \{2; 4; 6\}$, $a\simeq 0.65$ and $M\simeq 4\times 10^6\;M_{\odot}$.}\label{fig2}
\end{figure}

\begin{figure}
  \resizebox{\hsize}{!}{\includegraphics[angle=0]{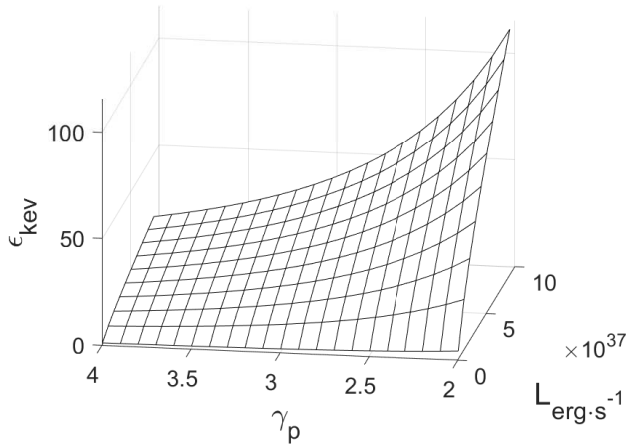}}
  \caption{Here we show the 3D plot of the photon's energy versus $\gamma_p$ and $L$.}\label{fig3}
\end{figure}

For the same set of parameters the synchrotron photon energy is given by
\begin{align}
  \epsilon\simeq 2.2\times 10^4\times\left(\frac{\gamma_{b}}{10^8}\right)^{5/2}\times\left(\frac{2}{\gamma_{p}}\right)^{5/2} \nonumber
    \\
    \times\left(\frac{L}{10^{37}ergs\; s^{-1}}\right)^{3/4}\times\left(\frac{r}{R_{lc}}\right)^{3/2}\textup{eV}.
\end{align}
In Fig. 1 we plot the dependence of photon energy$\epsilon$ on the bolometric luminosity, $L$ for the set of parameters: $\gamma_b = 10^8$, $\gamma_p = 2$, $a\simeq 0.65$ and $M\simeq 4\times 10^6\;M_{\odot}$. The emitted energy is in the hard $X$-ray band, and is a continuously increasing function of the bolometric luminosity - a natural consequence of the dependence of the magnetic field and the pitch angle on $L$ (see Eqs. \ref{B},\ref{pitchangle}). 

We have already mentioned that the synchrotron photon energy is sensitive to the values of the plasma $\gamma_p$. In Fig. 2, the plots of the emitted photon energy versus the bulk relativistic factor are shown for several values of $\gamma_p$ for the set of parameters: $L = 10^{37}$ergs s$^{-1}$, $\gamma_p = \{2; 3; 4\}$, $a\simeq 0.65$ and $M\simeq 4\times 10^6\;M_{\odot}$. We note that by increasing the plasma Lorentz factor by the factor of $2$, the photon energy  decreases significantly. A similar $3D$ plot is displayed in Fig. 3 where, apart from the $\gamma_p$ dependence, we  highlight how the photon energy depends on the bolometric luminosity. The set of parameters is the same as on Fig. 1, except $\gamma_p$ which varies over the range $[2-4]$.

This analysis shows that despite strong synchrotron losses in the nearby area of the central BH of the galaxy, the QLD maintains the continuous synchrotron emission process and might account for the generation of hard $X$-rays.

\section{Summary}

Perhaps the most significant contribution of this paper is to demonstrate how an instability induced QLD extends the times over which synchrotron emission may be maintained even for very high $\gamma$ electrons. For this purpose, the kinetic equation governing the particle distribution by pitch angles, was derived.

For physical parameters typical for the magnetosphere of SgrA$^*$, it had already  been found that the anomalous Doppler effect causes the cyclotron instability with the frequency of the order of $10^4$ Hz.  which means that this particular mode cannot escape a thick plasma ambient of the BH. It is precisely these waves that, via QLD,  "restore" the synchrotron process by diffusively injecting the perpendicular electron energy.

With QLD creating the right conditions for continuing synchrotron emission for typical $\gamma_p$s bolometric luminosities, one could safely expect photon energies up to $120$ keV lying in the hard $X$-ray spectral band.


\section*{Data Availability}

Data are available in the article and can be accessed via a DOI link.

\end{document}